\documentclass[aps,nofootinbib,preprint,superscriptaddress]{revtex4}%
\usepackage{hyperref}
\usepackage{amsmath}
\usepackage{amsfonts}
\usepackage{amssymb}
\usepackage{graphicx}
\usepackage{color}%
\providecommand{\U}[1]{\protect\rule{.1in}{.1in}}

\begin{document}
\title{Stringy scaling of n-point hard string scattering amplitudes}
\author{Sheng-Hong Lai}
\email{xgcj944137@gmail.com}
\affiliation{Department of Electrophysics, National Yang Ming Chiao Tung University,
Hsinchu, Taiwan, R.O.C.}
\author{Jen-Chi Lee}
\email{jcclee@cc.nctu.edu.tw}
\affiliation{Department of Electrophysics, National Yang Ming Chiao Tung University,
Hsinchu, Taiwan, R.O.C.}
\affiliation{Center for Theoretical and Computational Physics (CTCP), National Yang Ming
Chiao Tung University, Hsinchu, Taiwan, R.O.C.}
\author{Yi Yang}
\email{yiyang@mail.nctu.edu.tw}
\affiliation{Department of Electrophysics, National Yang Ming Chiao Tung University,
Hsinchu, Taiwan, R.O.C.}
\affiliation{Center for Theoretical and Computational Physics (CTCP), National Yang Ming
Chiao Tung University, Hsinchu, Taiwan, R.O.C.}
\date{\today}

\begin{abstract}
Motivated by the recent calculation of the $SL(K+3,%
\mathbb{C}
)$ symmetry of $n$-point Lauricella string scattering amplitudes (SSA) of open
bosonic string theory, we calculate ratios of the solvable infinite linear
relations of arbitrary $n$-point hard SSA ($HSSA$). We discover a general
\textit{stringy scaling} behavior for all $n$-point $HSSA$ to all string loop
orders. For the special case of $n=4$, the stringy scaling behavior reduces to
the infinite linear relations and constant ratios among $HSSA$ conjectured by
Gross \cite{Gross} and later corrected and calculated by the method of
decoupling of zero-norm states \cite{CHLTY2}.

\end{abstract}
\maketitle

\section{Introduction}

Symmetry principle was one of the most important discovery of the 20th century
physics. All four fundamental interactions of the universe were based on the
symmetry principles. In particular, the principles of the general coordinate
symmetry and the gauge symmetry have led to Einstein general relativity theory
and Yang-Mills gauge theory, respectively. In string theory, on the contrary,
the situation went the other way around. One is given a set of rules through
quantum consistency of the extended string to fix the forms of interactions or
vertices of the theory. Moreover, in contrast to up to the usual four-point
couplings in quantum field theory, e.g. QCD, in string theory, one encounters
arbitrary $n$-point couplings which correspond to the infinite number of
particles in the spectrum of string. As a result, it is crucial to identify
the huge symmetry structure of string theory and uses it to relate these
infinite number of couplings of particles with arbitrary higher masses and spins.

One key approach to uncover symmetry of string theory is to explicitly
calculate string scattering amplitudes (SSA). Recently, the author of
\cite{LLY2} calculated a subset of exact $4$-point SSA with three tachyons and
one arbitrary string states (Note that SSA of three tachyons and one arbitrary
string states with polarizations orthogonal to the scattering plane
\textit{vanish}.)%
\begin{equation}
\left\vert r_{n}^{T},r_{m}^{P},r_{l}^{L}\right\rangle =\prod_{n>0}\left(
\alpha_{-n}^{T}\right)  ^{r_{n}^{T}}\prod_{m>0}\left(  \alpha_{-m}^{P}\right)
^{r_{m}^{P}}\prod_{l>0}\left(  \alpha_{-l}^{L}\right)  ^{r_{l}^{L}}%
|0,k\rangle\label{state}%
\end{equation}
where $e^{P}=\frac{1}{M_{2}}(E_{2},\mathrm{k}_{2},0)=\frac{k_{2}}{M_{2}}$ is
the momentum polarization, $e^{L}=\frac{1}{M_{2}}(\mathrm{k}_{2},E_{2},0)$ is
the longitudinal polarization and $e^{T}=(0,0,1)$ is the transverse
polarization on the $\left(  2+1\right)  $-dimensional scattering plane, and
expressed them in terms of the $D$-type Lauricella functions \cite{LLY2}. In
addition to the mass level $M_{2}^{2}=2(N-1)$ with%
\begin{equation}
N=\sum_{\substack{n,m,l>0\\\{\text{ }r_{j}^{X}\neq0\}}}\left(  nr_{n}%
^{T}+mr_{m}^{P}+lr_{l}^{L}\right)  , \label{NN}%
\end{equation}
we define another important index $K$ for the state in Eq.(\ref{state})%
\begin{equation}
K=\sum_{\substack{n,m,l>0\\\{\text{ }r_{j}^{X}\neq0\}}}\left(  n+m+l\right)
\label{KKK}%
\end{equation}
where $X=\left(  T,P,L\right)  $ and we have put $r_{n}^{T}=r_{m}^{P}%
=r_{l}^{L}=1$ in Eq.(\ref{NN}) in the definition of $K$. Intuitively, $K$
counts the number of variaty of the $\alpha_{-j}^{X}$ oscillators in
Eq.(\ref{state}). For later use, we also define $k_{j}^{X}\equiv e^{X}\cdot
k_{j}$.

In addition, it was shown that these Lauricella SSA ($LSSA$) can be expressed
in terms of the basis functions in the infinite dimensional representation of
the $SL(K+3,%
\mathbb{C}
)$ group \cite{Group,slkc}. This is similar to the well-known spherical
harmonics function representation of the $SU(2)$ rotation group in quantum
mechanics. Moreover, it was further shown that there existed $K+2$ recurrence
relations of the $D$-type Lauricella functions. These recurrence relations can
be used to reproduce the Cartan subalgebra and simple root system of the
$SL(K+3,%
\mathbb{C}
)$ group with rank $K+2$. On the other hand, with the Cartan subalgebra and
the simple roots, one can easily deduce the whole Lie algebra of the $SL(K+3,%
\mathbb{C}
)$ group. So we have the following correspondences%
\begin{equation}
LSSA\Longleftrightarrow SL(K+3,%
\mathbb{C}
)\text{ symmetry }\Longleftrightarrow\text{ Recurrence relations of
Lauricella.} \label{lssa}%
\end{equation}
As a result, the $SL(K+3,%
\mathbb{C}
)$ group with its associated stringy Ward identities (recurrence relations)
can be used to solve \cite{solve} all the $LSSA$ and express them in terms of
one amplitude. See the recent review paper \cite{LSSA}.

One important application of Eq.(\ref{lssa}) in the hard string scattering
limit was to reproduce \cite{LLY2} infinite linear relations with constant
coefficients among all $4$-point hard SSA ($HSSA$) and solve the ratios among
them. This high energy symmetry of string theory \cite{GM,GM1} was first
conjectured by Gross \cite{Gross} and later corrected and proved
\cite{ChanLee,ChanLee2,CHLTY2,CHLTY1} by the method of decoupling of zero norm
states (ZNS) \cite{Lee,lee-Ov,LeePRL}. ZNS was also shown to carry $w_{\infty
}$ symmetry charges of $2D$ string ($c=1$, $2D$ quantum gravity)
\cite{ChungLee1}. See the review papers \cite{review,over}. More importantly,
since the decoupling of ZNS or stringy Ward identities persist to all string
loop orders \cite{loop}, it is conceivable that the infinite linear relations
obtained in the hard scattering limit for arbitrary mass $M^{2}=2(N-1)$ levels
for $4$-point SSA are also valid for all string loop amplitudes. On the other
hand, one notes that these linear relations are not shared by amplitudes of
quantum field theories which depend on polarizations of the scattering
particles \cite{spin}.

More recently, it was demonstrated \cite{short, long} that by using the string
on-shell recursion relation of SSA \cite{bcfw4,bcfw3,stringbcfw}, one can show
that \textit{all} $n$-point SSA can be expressed in terms of the Lauricella
functions. This result extends the previous exact $SL(K+3,%
\mathbb{C}
)$ symmetry of the $4$-point $LSSA$ of three tachyons and one arbitrary string
states in Eq.(\ref{lssa}) to the whole tree-level open bosonic string theory.
Motivated by this remarkable result \cite{short, long}, one is tempted to
believe that for the higher point ($n\geq5$) $LSSA$, there exist hard
scattering kinematics regimes for which the infinite linear relations persist
and the ratios can be solved accordingly.

In general, the explicit calculation of higher point ($n\geq5$) SSA is very
lengthy \cite{short, long}. In this letter, we will generalize the method of
decoupling of ZNS and calculate the solvable infinite linear relations and
ratios of \textit{arbitrary} $n$-point $HSSA$. The case of $n=4$ corresponds
to the previous Gross conjecture. In addition, we will use saddle point method
to re-calculate these infinite linear relations and ratios of a class of
$n$-point $HSSA$ to justify our results.

For each $n$ with $n\geq5$, we discover new ratios among $n$-point $HSSA$.
Moreover, we discover the reduction of both the number of kinematics variables
dependence on the ratios and the number of independent $HSSA$. These
\textit{stringy scaling} behaviors reminiscent of Bjorken scaling \cite{bs}
and the Callan-Gross relation \cite{cg} in deep inelastic scattering of
electron and proton in the quark-parton model of QCD are stringy phenomena
with energy much higher than the Planck energy ($\gg M_{Planck}^{2}%
\sim1/\alpha^{\prime}$), while the QCD scaling behavior is considered to be a
low energy ($\ll M_{Planck}^{2}$) string theory.

Nevertheless, there is another view to study hard scattering of QCD in string
theory, namely, string realization of QCD through gauge/string duality or
string on warped spacetime geometry \cite{ps,bt,mul}. In this picture, it was
shown that the hard string scattering is power-law and thus QCD-like. However,
since the scattering energy is of order of the string scale, it involves
physics beyond the supergravity approximation, and one encounters the
difficulty of solving string theory on curved spacetime. We believe that the
stringy scaling behavior discovered in this paper on flat spacetime background
and its possible extention to warped spacetime geometry may find application
on these issues, in particular, the realization of QCD scaling behavior in
string theory.

The stringy scaling behavior also reminds us of renormalization group (RG)
approach to the critical phenomena of phase transition \cite{Wilson} such as
ferromagnet, percolation, chaos, polymers and wind fetch. However, although
both theories are fixed-point theories ($D=26$ for the string case), the
stringy scaling or reduction of number of kinematics variables shows up only
in the hard scattering limit. See more discussion following Eq.(\ref{RG}).

On the other hand, there exists mechanism of (Spontaneously) broken symmetry
in the Landau-Wilson theory of phase transition where the low-temperature
phase has lost some symmetry. For the string theory side, there are two
different concepts of broken symmetry. The first one was provided by the
no-ghost theorem which says that the string spectrum is ghost-free provided
that the intercept $a=1$ and $D=26$ or $a\leq1$ and $D\leq25$ \cite{GSW}. For
the case of $D=26$ comparing to the latter case, there are enhanced $SL(K+3,%
\mathbb{C}
)$ symmetries due to the existence of the type II zero-norm states. The second
one was discovered more recently that the exact $SL(K+3,%
\mathbb{C}
)$ symmetry of $LSSA$ at $D=26$ was broken down to the $SL(5,%
\mathbb{C}
)$ symmetry in the Regge limit \cite{LSSA} and to the $SL(4,%
\mathbb{C}
)$ symmetry in the nonrelativistic limit \cite{LSSA}.

\section{Stringy scaling by decoupling of ZNS}

The use of method of decoupling of ZNS (or stringy Ward identities) to
calculate infinite linear relations for the $4$-point $HSSA$ was first adapted
in the "physical state basis" \cite{ChanLee,ChanLee2} with the Virasoro
constraints imposed. It was then extended to the general mass levels
$M^{2}=2(N-1)$ in the "oscillator state basis" without imposing the
constraints \cite{CHLTY1,CHLTY2}. We will generalize the method to the higher
point ($n\geq5$) $HSSA$ cases as will describe in this section.

In the old covariant first quantized open bosonic string spectrum, there are
two types of physical ZNS \cite{GSW}. While type I states have zero-norm at
any spacetime dimension, type II states have zero-norm only at $D=26$. The
starting point is to apply the $n$-point $l$-loop stringy on-shell Ward
identities \cite{JCLee}%
\begin{equation}
\left\langle V_{1}\chi V_{3}\cdots V_{n}\right\rangle _{l-loop}=0\label{w}%
\end{equation}
in the hard scattering limit. In Eq.(\ref{w}), $V_{j}$ can be any string
vertex and the second vertex $\chi$ is the vertex of a ZNS. In the hard
scattering limit of scattering processes \textit{on the scattering plane}, the
space part of momenta $k_{j}$ ( $j=3,4,\cdots,n$) form a closed $1$-chain with
$(n-2)$ sides due to momentum conservation and it can be shown that at each
fixed mass level $M^{2}=2(N-1)$ only states of the following form
\cite{CHLTY1,CHLTY2}
\begin{equation}
\left\vert N,2m,q\right\rangle =\left(  \alpha_{-1}^{\hat{T}}\right)
^{N-2m-2q}\left(  \alpha_{-1}^{L}\right)  ^{2m}\left(  \alpha_{-2}^{L}\right)
^{q}\left\vert 0;k\right\rangle \label{np}%
\end{equation}
are leading order in energy. In Eq.(\ref{np}) we have defined $e^{P}=\frac
{1}{M_{2}}(E_{2},\mathrm{k}_{2},\vec{0})=\frac{k_{2}}{M_{2}}$ the momentum
polarization, $e^{L}=\frac{1}{M_{2}}(\mathrm{k}_{2},E_{2},\vec{0})$ the
longitudinal polarization and the transverse polarization $e^{\hat{T}%
}=(0,0,\vec{\omega})$ where
\begin{equation}
\omega_{i}=\cos\theta_{i}\prod\limits_{\sigma=1}^{i-1}\sin\theta_{\sigma
}\text{ with }i=1,\cdots,r,\text{ }\theta_{r}=0\text{ }\label{ww}%
\end{equation}
are the solid angles in the transverse space spanned by $24$ transverse
directions $e^{T_{i}}$. Note that $\alpha_{-1}^{\hat{T}}=\alpha_{-1}\cdot
e^{\hat{T}}$ etc. Most importantly, to the leading hard scattering limit in
energy, one can identify $e^{P}\simeq$ $e^{L}$ \cite{ChanLee,ChanLee2}. This
is different from the Regge regime where $e^{P}\ncong$ $e^{L}$ \cite{bosonic}.
It is a straightforward calculation to show that, at each fixed mass level
$N$, the decoupling of ZNS leads to ratios among $n$-point $HSSA$
\cite{CHLTY1,CHLTY2}%
\begin{equation}
\frac{\mathcal{T}^{\left(  N,m,q\right)  }}{\mathcal{T}^{\left(  N,0,0\right)
}}=\frac{\left(  2m\right)  !}{m!}\left(  \frac{-1}{2M}\right)  ^{2m+q}%
\label{ratio5}%
\end{equation}
where $\mathcal{T}^{\left(  N,m,q\right)  }$ is the $n$-point $HSSA$ of any
string vertex $V_{j}$ with $j=1,3,\cdots n$, and $V_{2}$ is the high energy
state in Eq.(\ref{np}); while $\mathcal{T}^{\left(  N,0,0\right)  }$ is the
$n$-point $HSSA$ of any string vertex $V_{j}$ with $j=1,3,\cdots n$, and
$V_{2}$ is the leading Regge trajectory string state at mass level $N$. Note
that we have omitted the tensor indice of $V_{j}$ with $j=1,3,4$ and keep only
those of $V_{2}$ in $\mathcal{T}^{\left(  N,2m,q\right)  }$.

\subsection{Examples}

For $n=4$, Eq.(\ref{ratio5}) implies the ratios are independent of $1$
scattering angle $\varphi$ which reproduces Gross conjecture \cite{Gross}. For
$n=5$, Eq.(\ref{ratio5}) implies the ratios are independent of $3$ kinematics
variables ($2$ angles and $1$ fixed ratio of two infinite energies) or, for
simplicity, $3$ scattering "angles". For $n=6$, there are $5$ scattering
"angles". See the $r=1$ special case of Eq.(\ref{scale}) at the end of this
section. To illustrate the infinite linear relations, historically, the first
example of Eq.(\ref{ratio5}) was calculated for $n=4$ and $M^{2}=4$, $6$ in
the "physical state basis" with the Virasoro constraints imposed. At mass
level $M^{2}=4,$ there are two physical positive-norm states, one symmetric
spin-$3$ state and one anti-symmetric spin-$2$ state. The decoupling of ZNS
leads to $3$ linear relations among $4$ leading order $4$-point $HSSA$
\cite{JCLee,ChanLee,ChanLee2}%

\begin{align}
\mathcal{T}_{LLT}^{5\rightarrow3}+\mathcal{T}_{(LT)}^{3}  &  =0,\\
10\mathcal{T}_{LLT}^{5\rightarrow3}+\mathcal{T}_{TTT}^{3}+18\mathcal{T}%
_{(LT)}^{3}  &  =0,\\
\mathcal{T}_{LLT}^{5\rightarrow3}+\mathcal{T}_{TTT}^{3}+9\mathcal{T}%
_{[LT]}^{3}  &  =0,
\end{align}
which can be solved to get the ratios \cite{ChanLee,ChanLee2}%
\begin{equation}
\mathcal{T}_{TTT}:\mathcal{T}_{LLT}:\mathcal{T}_{(LT)}:\mathcal{T}%
_{[LT]}=8:1:-1:-1. \label{h}%
\end{equation}
Eq.(\ref{h}) calculated in the "physical state basis" is consistent with the
result of Eq.(\ref{ratio5}) calculated in the "oscillator state basis" in
Eq.(\ref{np}). It was also demonstrated that the next to leading order
amplitudes are in general not proportional to each other
\cite{ChanLee,ChanLee2}.

In general scattering processes out of the scattering plane, the general high
energy states at each fixed mass level $M^{2}=2(N-1)$ can be written as%
\begin{equation}
\left\vert \left\{  p_{i}\right\}  ,2m,2q\right\rangle =\left(  \alpha
_{-1}^{T_{1}}\right)  ^{N+p_{1}}\left(  \alpha_{-1}^{T_{2}}\right)  ^{p_{2}%
}\cdots\left(  \alpha_{-1}^{T_{r}}\right)  ^{p_{r}}\left(  \alpha_{-1}%
^{L}\right)  ^{2m}\left(  \alpha_{-2}^{L}\right)  ^{q}\left\vert
0;k\right\rangle \label{pr}%
\end{equation}
where $\sum_{i=1}^{r}p_{i}=-2(m+q)$ with\ $r\leq24$. With $\left(  \alpha
_{-1}^{T_{i}}\right)  =\left(  \alpha_{-1}^{\hat{T}}\right)  \omega_{i}$, we
easily obtain%
\begin{align}
&  \left(  \alpha_{-1}^{T_{1}}\right)  ^{N+p_{1}}\left(  \alpha_{-1}^{T_{2}%
}\right)  ^{p_{2}}\cdots\left(  \alpha_{-1}^{T_{r}}\right)  ^{p_{r}}\left(
\alpha_{-1}^{L}\right)  ^{2m}\left(  \alpha_{-2}^{L}\right)  ^{q}\left\vert
0;k\right\rangle \nonumber\\
&  =\left(  \omega_{1}^{N}\prod_{i=1}^{r}\omega_{i}^{p_{i}}\right)  \left(
\alpha_{-1}^{\hat{T}}\right)  ^{N-2m-2q}\left(  \alpha_{-1}^{L}\right)
^{2m}\left(  \alpha_{-2}^{L}\right)  ^{q}\left\vert 0;k\right\rangle ,
\end{align}
which leads to the ratios of $n$-point $HSSA$%
\begin{equation}
\frac{\mathcal{T}^{\left(  \left\{  p_{i}\right\}  ,2m,2q\right)  }%
}{\mathcal{T}^{\left(  \left\{  0_{i}\right\}  ,0,0\right)  }}=\frac{\left(
2m\right)  !}{m!}\left(  \frac{-1}{2M}\right)  ^{2m+q}\prod_{i=1}^{r}%
\omega_{i}^{p_{i}} \label{ratiog}%
\end{equation}
where $\mathcal{T}^{\left(  \left\{  0_{i}\right\}  ,0,0\right)  }$ is the
$HSSA$ of leading Regge trajectory state at mass level $M^{2}=2(N-1)$. These
ratios are valid to all string loop orders.

We are now ready to discuss the stringy scaling behavior of these $n$-point
$HSSA$. First, we would like to count the number of independent kinematics
variables of the $n$-point $HSSA$. For the simple case with $n=4$ and $r=1$ in
Eq.(\ref{pr}) or Eq.(\ref{np}), one has two variables, $s$ and $t$, or energy
$E$ and the scattering angle $\varphi$, and Eq.(\ref{ratio5}) means that all
$HSSA$ at each fixed mass level are proportional to each other and the ratios
are independent of the scattering angle $\varphi$. For the general $n$-point
$HSSA$ with $r\leq24$ in Eq.(\ref{pr}), we have $k_{j}$ vector with
$j=1,\cdots,n$ and $k_{j}$ $\in R^{d-1,1}$. We can count the number of
independent kinematics variables to be $n\left(  d-1\right)  -\frac{d\left(
d+1\right)  }{2}$.

In the hard scattering limit, for the kinematics parameter space $\mathcal{M}$
defined by
\begin{equation}
\omega_{j}\left(  \text{kinematics parameters with }E\rightarrow\infty\right)
=\text{fixed constant \ }(j=2,\cdots,r)\text{,}%
\end{equation}
we can count the dimension of $\mathcal{M}$ to be
\begin{equation}
\text{dim}\mathcal{M}\text{ }\mathcal{=}\text{ }n\left(  d-1\right)
-\frac{d\left(  d+1\right)  }{2}-1-\left(  r-1\right)  =\frac{\left(
r+1\right)  \left(  2n-r-6\right)  }{2}. \label{m}%
\end{equation}
where $r=d-2$ is the number of transverse directions $e^{T_{i}}$. In sum, the
ratios among $n$-point $HSSA$ with $r\leq24$ calculated in Eq.(\ref{ratiog})
are constants and independent of the scattering "angles" in $\mathcal{M}$. For
$n=6$ and $r=3$, as an example, $\mathcal{M}$ is defined by
\begin{equation}
\theta_{j}\left(  8\text{ kinematics parameters}\right)  =\text{fixed
constant, \ }j=1,2, \label{gg}%
\end{equation}
and we have $\dim\mathcal{M=}$ $6$. For this case, the ratios%
\begin{equation}
\frac{\mathcal{T}^{\left(  \left\{  p_{1},p_{2},p_{3}\right\}  ,m,q\right)  }%
}{\mathcal{T}^{\left(  \left\{  0,0,0\right\}  ,0,0\right)  }}=\frac{\left(
2m\right)  !}{m!}\left(  \frac{-1}{2M}\right)  ^{2m+q}\left(  \cos\theta
_{1}\right)  ^{p_{1}}\left(  \sin\theta_{1}\cos\theta_{2}\right)  ^{p_{2}%
}\left(  \sin\theta_{1}\sin\theta_{2}\right)  ^{p_{3}} \label{16}%
\end{equation}
are independent of kinematics parameters in the space $\mathcal{M}$. For
example, for say $\theta_{1}=\frac{\pi}{4}$ and $\theta_{2}=\frac{\pi}{6}$, we
get the ratios among $6$-point $HSSA$%
\begin{equation}
\frac{\mathcal{T}^{\left(  \left\{  p_{1},p_{2},p_{3}\right\}  ,m,q\right)  }%
}{\mathcal{T}^{\left(  \left\{  0,0,0\right\}  ,0,0\right)  }}=\left(
-\frac{1}{M}\right)  ^{2m+q}(2m-1)!!\left(  \frac{1}{2}\right)  ^{p_{2}+p_{3}%
}\left(  \sqrt{3}\right)  ^{p_{3}}\text{.}%
\end{equation}

We conclude that, in the hard scattering limit, the number of scattering
"angles" dependence on ratios of $n$-point $HSSA$ with $r\leq24$ reduces by
dim$M$. For a given ($n,r$), we list some numbers of dim$\mathcal{M}$
calculated in Eq.(\ref{m}) as following
\begin{equation}%
\begin{tabular}
[c]{lllll}%
$\text{dim}\mathcal{M}$ & $r=1$ & $r=2$ & $r=3$ & $r=4$\\
$n=4$ & $1$ &  &  & \\
$n=5$ & $3$ & $3$ &  & \\
$n=6$ & $5$ & $6$ & $6$ & \\
$n=7$ & $7$ & $9$ & $10$ & $10$%
\end{tabular}
\ \ \ \ \ \ \ \ . \label{scale}%
\end{equation}
We see from Eq.(\ref{ratiog}) that form of the ratios calculated for a given
$r$ are valid for arbitrary $n$-point $HSSA$ and to all string loop orders.
Note that for a given $n$,$\ \max(r)=n-3$, we have dim$\mathcal{M}\left(
r=n-4\right)  =$ dim$\mathcal{M}\left(  r=n-3\right)  =\frac{\left(
n-3\right)  \left(  n-2\right)  }{2}.$

The reduction of both the number of scattering "angles" dependence on ratios
and independent SSA in the hard scattering limit suggests a very
interesting\ phenomenon of \textit{stringy scaling} behavior.

\section{String amplitudes in the hard limit}

To do a consistent check of the ratios calculated in Eq.(\ref{ratiog}) by the
decoupling of ZNS, in this section, we will use saddle point method to
calculate the $n$-point $HSSA$. Since the ratios in Eq.(\ref{ratiog}) are
independent of the choices of $V_{j}$ with $j=1,3,\cdots n$, we choose them to
be tachyons and $V_{2}$ is chosen to be Eq.(\ref{pr}). On the other hand,
since the ratios are independent of the loop order, we choose to calculate
$l=0$ loop. In the hard scattering limit, $p=E\rightarrow\infty$, we define
the $26$-dimensional momenta in the CM frame to be
\begin{align}
k_{1} &  =\left(  E,-E,0^{r}\right)  ,\nonumber\\
k_{2} &  =\left(  E,+E,0^{r}\right)  ,\nonumber\\
&  \vdots\nonumber\\
k_{j} &  =\left(  -q_{j},-q_{j}\Omega_{1}^{j},-q_{j}\Omega_{2}^{j}%
,\cdots,-q_{j}\Omega_{r}^{j},-q_{j}\Omega_{r+1}^{j}\right)  \label{k2}%
\end{align}
where $j=3,4,\cdots,n$, and%
\begin{equation}
\Omega_{i}^{j}=\cos\phi_{i}^{j}\prod\limits_{\sigma=1}^{i-1}\sin\phi_{\sigma
}^{j}\text{ with }\phi_{j-1}^{j}=0,\text{ }\phi_{i>r}^{j}=0\text{ and }%
r\leq\min\left\{  n-3,24\right\}  \label{kk}%
\end{equation}
are the solid angles in the $\left(  j-2\right)  $-dimensional spherical space
with $\sum_{i=1}^{j-2}\left(  \Omega_{i}^{j}\right)  ^{2}=1$. In
Eq.(\ref{k2}), $0^{r}$ denotes the $r$-dimensional null vector. The condition
$\phi_{j-1}^{j}=0$ in Eq.(\ref{kk}) was chosen to fix the frame by using the
rotational symmetry. Note that the space part of $k_{j}$ ( $j=3,4,\cdots,n$)
form a closed $1$-chain with $(n-2)$ sides in the $25$-dimensional space due
to momentum conservation. In the hard scattering limit $E\rightarrow\infty$,
the independent kinematics variables counted in the last section can be chosen
to be some $\varphi_{i}^{j}$ and some fixed ratios of infinite $q_{j}$ defined
in Eq.(\ref{k2}). We consider the $n$-point $HSSA$ in $26D$ spacetime with
$n-1$ tachyons and $1$ state defined in Eq.(\ref{pr}). With the change of
variables $z_{i}=\frac{x_{i}}{x_{i+1}}$ or $x_{i}=z_{i}\cdots z_{n-2}$, the
$n$-point $HSSA$ can be calculated to be%
\begin{equation}
\mathcal{T}^{\left(  \left\{  p_{i}\right\}  ,m,q\right)  }=\int_{0}%
^{1}dx_{n-2}\cdots\text{ }\int_{0}^{x_{4}}dx_{3}\int_{0}^{x_{3}}dx_{2}%
ue^{-Kf}=\left(  \prod_{i=3}^{n-2}\int_{0}^{1}dz_{i}\text{ }z_{i}%
^{i-2-N}\right)  \int_{0}^{1}dz_{2}ue^{-Kf}\label{int}%
\end{equation}
where%
\begin{align}
f &  =-\underset{i<j}{\sum}\frac{k_{i}\cdot k_{j}}{K}\ln\left(  x_{j}%
-x_{i}\right)  \\
&  =-\underset{i<j}{\sum}\frac{k_{i}\cdot k_{j}}{K}\left[  \ln(z_{j}\cdots
z_{n-2})+\ln\left(  1-z_{i}\cdots z_{j-1}\right)  \right]  ,\text{ }%
K=-k_{1}\cdot k_{2},\\
u &  =\left(  k^{T_{1}}\right)  ^{N+p_{1}}\left(  k^{T_{2}}\right)  ^{p_{2}%
}\cdots\left(  k^{T_{r}}\right)  ^{p_{r}}\left(  k^{L}\right)  ^{2m}\left(
k^{\prime L}\right)  ^{q}.(k^{\prime L}=\frac{\partial k^{L}}{\partial x_{2}%
})\label{uu}%
\end{align}
In Eq.(\ref{uu}), we have defined%
\begin{equation}
k=-\sum_{i\neq2,n}\frac{k_{i}}{\left(  z_{i}\cdots z_{n-2}-z_{2}z_{3}\cdots
z_{n-2}\right)  }=\sum_{i\neq2,n}\frac{k_{i}}{x_{i}-x_{2}}\label{kp}%
\end{equation}
and $k^{X}$ is the momentum $k$ projected on the $X$ polarization. We also
define $k_{\perp}=\left\vert k_{\perp}\right\vert \sum_{i=1}^{r}e^{T_{i}%
}\omega_{i}=\left\vert k_{\perp}\right\vert e^{\hat{T}}$. We will show that
the new defined vector $k$ is the key quantity to connect the stringy scaling
variables $\omega_{i}$ or $\theta_{i}$ introduced in Eq.(\ref{ww}) and the
kinematic variables $\Omega_{i}^{j}$ or $\varphi_{i}^{j}$ introduced in the CM
frame defined in Eq.(\ref{kk}).

To do the integration in Eq.(\ref{int}), we define the saddle point $\tilde
{z}_{i}=\left(  \tilde{z}_{2},\cdots,\tilde{z}_{n-2}\right)  $ to be the
solution of%
\begin{equation}
\frac{\partial f}{\partial z_{2}}=0,\cdots,\frac{\partial f}{\partial z_{n-2}%
}=0. \label{f2}%
\end{equation}
Note that Eq.(\ref{f2}) implies%
\begin{align}
\tilde{k}^{L}  &  =\frac{\tilde{k}\cdot k_{2}}{M}=\frac{k_{12}}{M}\left.
\frac{\partial f}{\partial x_{2}}\right\vert _{z_{i}=\tilde{z}_{i}}\nonumber\\
&  =\frac{k_{12}}{M}\left.  \frac{\partial z_{j}}{\partial x_{2}}%
\frac{\partial f}{\partial z_{j}}\right\vert _{z_{i}=\tilde{z}_{i}}=0,\text{
}\left\vert \tilde{k}\right\vert =\left\vert \tilde{k}_{\perp}\right\vert
\text{. } \label{kkk}%
\end{align}
To proceed, let us introduce the following key identity%
\begin{equation}
\tilde{k}^{2}+2M\tilde{k}^{\prime L}=0, \label{iden}%
\end{equation}
which can be interpreted as a mathematical identity of the geometric
parameters of a closed $1$-chain with $n-2$ sides. For $n=4$, one can easily
solve the saddle point $\tilde{z}_{2}=\sec^{2}\frac{\varphi}{2}$ to prove
Eq.(\ref{iden}) analytically. For the cases of $n=5$ and $6$, we are able to
prove the identity numerically. Indeed, we will see that the identity is a
result of the calculation of decoupling of ZNS in Eq.(\ref{ratiog}).
Eq.(\ref{iden}) is crucial to show the stringy scaling behavior of $HSSA$. On
the other hand, Eq.(\ref{iden}) can be used to express $\omega_{i}$ (or
$\theta_{i}$) defined in Eq.(\ref{ww}) in terms of independent kinematics
variables defined in Eq.(\ref{k2}) if one can analytically solve the saddle
point $\tilde{z}_{i}=\left(  \tilde{z}_{2},\cdots,\tilde{z}_{n-2}\right)  $.
Unfortunately, it turns out to be nontrivial except for the case of $n=4$. For
$n=6$ and $r=3$, see Eq.(\ref{gg}) as an example, we can only formally express
$\theta_{i}$ in terms of kinematics variables defined in Eq.(\ref{k2}) as%
\begin{equation}
\theta_{1}=\arctan\frac{\sqrt{\left(  \tilde{k}_{\bot}^{T_{2}}\right)
^{2}+\left(  \tilde{k}_{\bot}^{T_{3}}\right)  ^{2}}}{\tilde{k}_{\perp}^{T_{1}%
}}\text{, }\theta_{2}=\arctan\frac{\tilde{k}_{\perp}^{T_{3}}}{\tilde{k}%
_{\perp}^{T_{2}}}. \label{RG}%
\end{equation}

It is important to note from Eq.(\ref{16}) that the ratios of $6$-point $HSSA$
depends only on $2$ variables $\theta_{1}$ and $\theta_{2}$ instead of $8$.
This stringy scaling behavior is reminiscent of the model-independent
relations or \textit{scaling relations }among critical exponents through RG
analysis of Widom's hypothesis in statistical mechanics where a set of $6$
exponents $\alpha$, $\beta$, $\gamma$, $\delta$, $\nu$ and $\eta$ are related
by $4$ scaling relations, and the number of independent exponents reduces from
$6$ to $2$.

In view of Eq.(\ref{uu}) and Eq.(\ref{kkk}), all up to $(2m)$-order
differentiations of $u$ function in Eq.(\ref{uu}) at the saddle point vanish
except%
\begin{equation}
\left.  \frac{\partial^{2m}u}{\partial z_{2}^{2m}}\right\vert _{z_{i}%
=\tilde{z}_{i}}=\frac{\left(  2m\right)  !}{\left(  -2M\right)  ^{2m+q}%
}\left\vert \tilde{k}\right\vert ^{N+2m}\omega_{1}^{N}\prod_{i=1}^{r}%
\omega_{i}^{p_{i}} \label{2m}%
\end{equation}
where we have used the \textit{identity} Eq.(\ref{iden}). With the above
inputs, we can calculate the $n$-point $HSSA$%
\begin{align}
\mathcal{T}^{\left(  \left\{  p_{i}\right\}  ,2m,2q\right)  }  &
\simeq\left(  \prod_{i=3}^{n-2}\int_{0}^{1}dz_{i}\text{ }z_{i}^{i-2-N}\right)
\int_{0}^{1}dz_{2}\left(  \frac{\partial^{2m}\tilde{u}}{\partial z_{2}^{2m}%
}\frac{\left(  z_{2}-\tilde{z}_{2}\right)  ^{2m}}{\left(  2m\right)
!}\right)  e^{-Kf}\nonumber\\
&  \simeq\frac{1}{\left(  2m\right)  !}\frac{\partial^{2m}\tilde{u}}{\partial
z_{2}^{2m}}\left(  \prod_{i=3}^{n-2}\tilde{z}_{i}^{i-2-N}\right)  \int%
_{0}^{\infty}dz_{2}\left(  z_{2}-\tilde{z}_{2}\right)  ^{2m}e^{-Kf\left(
z_{2}\right)  }\nonumber\\
&  =2\sqrt{\pi}e^{-K\tilde{f}}\left\vert \tilde{k}\right\vert ^{N-1}\left(
\prod_{i=3}^{n-2}\tilde{z}_{i}^{i-2-N}\right)  \omega_{1}^{N}\frac{\left(
2m\right)  !}{m!}\left(  \frac{-1}{2M}\right)  ^{2m+q}\prod_{i=1}^{r}%
\omega_{i}^{p_{i}} \label{main}%
\end{align}
where $f\left(  z_{2}\right)  =f\left(  z_{2},\tilde{z}_{3},\cdots,\tilde
{z}_{n-2}\right)  $. Eq.(\ref{main}) gives the ratios of $n$-point $HSSA$ in
Eq.(\ref{ratiog}). If we confine the $n$-point $HSSA$ on a scattering plane,
i.e. $r=1$ and $\omega_{1}=1$, the ratios in Eq.(\ref{ratiog}) reduces to the
ratios of $n$-point $HSSA$ in Eq.(\ref{ratio5}). For the special $n=4$ case,
one reproduces the ratios obtained in \cite{CHLTY1,CHLTY2} for the Gross
conjecture on $4$-point $HSSA$. Thus Gross conjucture is just a special
$4$-point case of ratios of $n$-point $HSSA$ with $r=1$ in Eq.(\ref{ratio5}).

To illustrate the calculation in Eq.(\ref{main}), the ratios of $n$-point
$HSSA$ for the case of $r=2$ can be calculated to be%
\begin{align}
\frac{\mathcal{T}^{\left(  p_{1},p_{2},m,q\right)  }}{\mathcal{T}^{\left(
N,0,0,0\right)  }}  &  =\frac{\left(  2m\right)  !}{m!}\left(  \frac{-1}%
{2M}\right)  ^{2m+q}\frac{\left(  2K\tilde{f}_{22}\right)  ^{m+q}}{\left(
\sum_{i\neq2,n}\frac{k_{i}^{T_{1}}}{\tilde{x}_{i}-\tilde{x}_{2}}\right)
^{2m+2q+p_{2}}\left(  \sum_{i\neq2,n}\frac{k_{i}^{T_{2}}}{\tilde{x}_{i}%
-\tilde{x}_{2}}\right)  ^{-p_{2}}}\nonumber\\
&  =\frac{\left(  2m\right)  !}{m!}\left(  \frac{-1}{2M}\right)  ^{2m+q}%
\frac{\left(  \frac{\sum_{i\neq2,n}\frac{k_{i}^{T_{2}}}{\tilde{x}_{i}%
-\tilde{x}_{2}}}{\sum_{i\neq2,n}\frac{k_{i}^{T_{1}}}{\tilde{x}_{i}-\tilde
{x}_{2}}}\right)  ^{p_{2}}}{\left(  \frac{\sum_{i\neq2,n}\frac{k_{i}^{T_{1}}%
}{\tilde{x}_{i}-\tilde{x}_{2}}}{\sqrt{2K\tilde{f}_{22}}}\right)  ^{2m+2q}}
\label{a2}%
\end{align}
where $\tilde{f}_{22}$ is defined in the following
\begin{equation}
f_{2}\equiv\frac{\partial f}{\partial z_{2}}\text{, }f_{22}\equiv
\frac{\partial^{2}f}{\partial z_{2}^{2}}\text{, }\tilde{f}=f\left(  \tilde
{z}_{2},\cdots,\tilde{z}_{n-2}\right)  \text{, }\tilde{f}_{22}=\left.
\frac{\partial^{2}f}{\partial z_{2}^{2}}\right\vert _{\left(  \tilde{z}%
_{2},\cdots,\tilde{z}_{n-2}\right)  }.
\end{equation}
On the other hand, the decoupling of ZNS of Eq.(\ref{ratiog}) gives%
\begin{equation}
\frac{\mathcal{T}^{\left(  p_{1},p_{2},m,q\right)  }}{\mathcal{T}^{\left(
N,0,0,0\right)  }}=\frac{\left(  2m\right)  !}{m!}\left(  \frac{-1}%
{2M}\right)  ^{2m+q}\omega_{1}^{p_{1}}\omega_{2}^{p_{2}}=\frac{\left(
2m\right)  !}{m!}\left(  \frac{-1}{2M}\right)  ^{2m+q}\frac{(\tan\theta
_{1})^{p_{2}}}{(\cos\theta_{1})^{2m+2q}}. \label{b2}%
\end{equation}
The saddle point calculation in Eq.(\ref{a2}) and the ZNS calculation in
Eq.(\ref{b2}) can be identified for any $p_{2}$, $m$ and $q$ if%
\begin{equation}
\left(  \sum_{i\neq2,n}\frac{k_{i}^{T_{1}}}{\tilde{x}_{i}-\tilde{x}_{2}%
}\right)  =\sqrt{2K\tilde{f}_{22}}\cos\theta_{1}\text{, }\left(  \sum
_{i\neq2,n}\frac{k_{i}^{T_{2}}}{\tilde{x}_{i}-\tilde{x}_{2}}\right)
=\sqrt{2K\tilde{f}_{22}}\sin\theta_{1},
\end{equation}
which implies the identity%
\begin{equation}
\left(  \sum_{i\neq2,n}\frac{k_{i}^{T_{1}}}{\tilde{x}_{i}-\tilde{x}_{2}%
}\right)  ^{2}+\left(  \sum_{i\neq2,n}\frac{k_{i}^{T_{2}}}{\tilde{x}%
_{i}-\tilde{x}_{2}}\right)  ^{2}=2K\tilde{f}_{22}. \label{qq2}%
\end{equation}
The identity derived in Eq.(\ref{qq2}) is the $r=2$ special case of the
identity in Eq.(\ref{iden}).

\section{Conclusion}

In this paper, we extend the previous calculation of ratios among $HSSA$ for
$n=4$ with $r=1$ in Eq.(\ref{scale}), which were conjectured by Gross
\cite{Gross} and calculated in \cite{ChanLee,ChanLee2,CHLTY2,CHLTY1}, to
obtain new ratios for general $n$-point $HSSA$ with $r\leq24$. Moreover, we
discover the reduction of both the number of kinematics variables dependence
on the ratios and the number of independent SSA in the hard string scattering
limit. These stringy scaling behaviors are reminiscent of deep inelastic
scattering of electron and proton where the two structure functions
$W_{1}(Q^{2},\nu)$ and $W_{2}(Q^{2},\nu)$ scale, and become not functions of
kinematics variables $Q^{2}$ and $\nu$ independently but only of their ratio
$Q^{2}/\nu$. That is, the structure functions scale as \cite{bs}%
\begin{equation}
MW_{1}(Q^{2},\nu)\rightarrow F_{1}(x),\text{ \ \ }\nu W_{2}(Q^{2}%
,\nu)\rightarrow F_{2}(x)
\end{equation}
where $x$ is the Bjorken variable and $M$ is the proton mass. Moreover, due to
the spin-$\frac{1}{2}$ assumption of quark, Callan and Gross derived the
relation \cite{cg}%
\begin{equation}
2xF_{1}(x)=F_{2}(x).
\end{equation}

These scaling behaviors in the hard scattering limit of quark-parton model in
QCD seems to persist in some way in the $HSSA$ of string theory although the
latter are stringy phenomena with energy much higher than the Planck energy,
while the QCD scaling behavior is considered to be a low energy ($\ll
M_{Planck}^{2}$) string theory. Nevertheless, there seems to be interesting
similarities between the two scaling behaviors. For given $n$-point $HSSA$
with $r\leq24$, the number of independent kinematics variables on the ratios
of $HSSA$ reduced by dim$\mathcal{M}$ as was calculated in Eq.(\ref{scale}).
Moreover, the number of independent $n$-point $HSSA$ for a given $r$ also
reduces for each fixed mass level as can be seen in Eq.(\ref{ratiog}).

We believe that, comparing to hard QCD scaling, hard string theory in general
has not been well studied yet in the literature \cite{oleg}. More new
phenomena of stringy scaling proposed in this letter remain to be uncovered.

This work is supported in part by the Ministry of Science and Technology
(MoST) and S.T. Yau center of National Yang Ming Chiao Tung University (NYCU), Taiwan.


\end{document}